\documentclass[preprint,12pt]{elsarticle}

\usepackage{amssymb}
\usepackage{lscape}
\usepackage{soul}
\usepackage{hyperref}

 \usepackage{lineno}

\biboptions{sort&compress}

\begin{document}

\begin{frontmatter}



\title{Study of cosmogenic activation above ground for the DarkSide-20k experiment}

\author	[UCDavis] {E.~Aaron} 
\author	[AQGSSI]		{P.~Agnes} 
\author	[AstroCeNT]			{I.~Ahmad}
\author	[CTINFN,CTUNI]		{S.~Albergo}
\author	[USP]			{I.~F.~M.~Albuquerque}
\author	[PNNLaddress]			{T.~Alexander}
\author	[Augustana]			{A.~K.~Alton}
\author	[TRIUMFaddress]			{P.~Amaudruz}
\author	[CAINFN]	{M.~Atzori Corona} 
\author	[USP]			{M.~Ave}
% ...

\author{(The DarkSide-20k Collaboration)\footnote{e-mail: ds-ed@lngs.infn.it}}

\address[UCDavis]{Department of Physics, University of California, Davis, CA 95616, USA}
\address[AQGSSI]{Gran Sasso Science Institute, L'Aquila 67100, Italy}
\address[AstroCeNT]{AstroCeNT, Nicolaus Copernicus Astronomical Center of the Polish Academy of Sciences, 00-614 Warsaw, Poland}
\address[CTINFN]{INFN Catania, Catania 95121, Italy}
\address[CTUNI]{Universit\`a of Catania, Catania 95124, Italy}
\address[USP]{Instituto de F\'isica, Universidade de S\~ao Paulo, S\~ao Paulo 05508-090, Brazil}
\address[PNNLaddress]{Pacific Northwest National Laboratory, Richland, WA 99352, USA}
\address[Augustana]{Physics Department, Augustana University, Sioux Falls, SD 57197, USA}
\address[TRIUMFaddress]{TRIUMF, 4004 Wesbrook Mall, Vancouver, BC V6T 2A3, Canada}
\address[CAINFN]{INFN Cagliari, Cagliari 09042, Italy}
% ...

\begin{abstract}
The activation of materials due to exposure to cosmic rays may become an important background source for experiments investigating rare event phenomena. DarkSide-20k, currently under construction at the Laboratori Nazionali del Gran Sasso, is a direct detection experiment for galactic dark matter particles, using a two-phase liquid-argon Time Projection Chamber (TPC) filled with 49.7~tonnes (active mass) of Underground Argon (UAr) depleted in $^{39}$Ar. Despite the outstanding capability of discriminating $\gamma$/$\beta$ background in argon TPCs, this background must be considered because of induced dead time or accidental coincidences mimicking dark-matter signals and it is relevant for low-threshold electron-counting measurements. Here, the cosmogenic activity of relevant long-lived radioisotopes induced in the experiment has been estimated to set requirements and procedures during preparation of the experiment and to check that it is not dominant over primordial radioactivity; particular attention has been paid to the activation of the 120~t of UAr used in DarkSide-20k.  Expected exposures above ground and production rates, either measured or calculated, have been considered in detail. From the simulated counting rates in the detector due to cosmogenic isotopes, it is concluded that activation in copper and stainless steel is not problematic. The activity of $^{39}$Ar induced during extraction, purification and transport on surface is evaluated to be 2.8\% of the activity measured in UAr by DarkSide-50 experiment, which used the same underground source, and thus considered acceptable. Other isotopes in the UAr such as $^{37}$Ar and $^{3}$H are shown not to be relevant due to short half-life and assumed purification methods.
\end{abstract}


\end{frontmatter}


\section{Introduction}
\label{secintro}

Great efforts have been devoted worldwide to unravel the nature of dark matter \cite{historydm} which is expected to fill our galaxy. One strategy  is to search for Weakly Interacting Massive Particles (WIMPs) by direct detection via WIMP-nucleus elastic scattering using of different kinds of sensitive radiation detectors~\cite{schumann2019,appecdm}. Noble elements like xenon and argon are ideal targets because the material is easily purified and detectors can be scaled in mass for high senstivity. \cite{pandax4t,xenonnt,lz,deap3600,darkside50,darkside50le1,darkside50le2}.

The expected counting rate from the interaction of WIMPs is extremely low, requiring ultra-low background conditions. This is achieved by operating in deep underground locations, using active and passive shielding, carefully selecting radiopure materials, and developing background-rejection methods in analysis~\cite{heusser,formaggio}. In this context, long-lived radioactive isotopes induced in the materials of the experiment by the exposure to cosmic rays during fabrication, transport and storage can be as relevant as residual contamination from primordial nuclides. In principle, cosmogenic activation can be kept under control by minimizing exposure on the surface and storing materials underground, avoiding flights, and even using shielding against the hadronic component of cosmic rays. It would be desirable to have reliable estimates of activation yields to assess the real danger of exposing materials to cosmic rays. Direct assay measurements of exposed materials, in very low background conditions, and calculations of production rates and yields, following different approaches, have been made for several materials in the context of dark matter, neutrinoless 2$\beta$ decay, and solar neutrino experiments \cite{cebrian,cebrianuniverse}. Results have been calculated for detector media such as germanium \cite{barabanov,mei,elliot2010,cebrianap,edelweisscos,tritiumpaper,cdmslitecos,cdexcos,yan}, silicon \cite{saldanhasi}, NaI~\cite{cebrianjcap,naicos,tritiumpaper,cosinecos,saldanhaNaI}, tellurium and TeO$_{2}$~\cite{te,telozza,wang}, xenon~\cite{schumann,mei2016,xeAr37}, argon~\cite{tritiumpaper,saldanha,zhangmei} and molybdate~\cite{chen} as well as for copper \cite{cebrianap,schumann,mei2016,coppers,she}, lead \cite{giuseppe} or stainless steel~\cite{mei2016,coppers}.

Liquid Argon (LAr) provides an outstanding Pulse Shape Discrimination (PSD) power to separate electron recoils (ER) from nuclear recoil (NR) events, as shown by the single-phase LAr detector DEAP-3600 \cite{deap3600}. Dual-phase Time Projection Chambers (TPCs) have additional capabilities like excellent spatial resolution. The DarkSide-50 experiment at the Laboratori Nazionali del Gran Sasso (LNGS) in Italy followed this approach using Underground Argon (UAr) (depleted of $^{39}$Ar by a factor 1400$\pm$200 with respect to the Atmospheric Argon (AAr) activity of $\sim$1~Bq/kg) \cite{darkside50,darkside50le1,darkside50le2}. Despite these excellent background discrimination capabilities, acceptance losses (via ER + NR pile-up in the TPC or accidental coincidence between the Veto and TPC signals that mimic the neutron capture signature) can be produced by $\gamma$ or $\beta$ emitters in the set-up; therefore, these background sources must be carefully considered too. The goal of this work is, considering exposure on the Earth's surface under realistic conditions, to quantify the yields of cosmogenic activation of detector materials and the effect on the expected counting rates of the DarkSide-20k detector; the results will be compared with those from other radioactive backgrounds like $^{39}$Ar. This allows requirements and procedures during the preparation and commissioning of the experiment to be set. The study has been carried out for UAr as well as for copper, and stainless steel, since the use of large quantities of these materials is foreseen in different components, according to the design of DarkSide-20k. The paper is structured as follows: the DarkSide-20k project is presented in Sec.~\ref{secdsproject}; the methodology applied to quantify cosmogenic activities is described in Sec.~\ref{cal}, showing the obtained results for different materials in Secs.~\ref{secCuSt} and \ref{secAr}; the counting rates expected from these activities are discussed in Sec.~\ref{secRates}, before summarizing conclusions in Sec.~\ref{con}.

\section{The GADMC and the DarkSide-20k detector}
\label{secdsproject}

The Global Argon Dark Matter Collaboration (GADMC) has been established to push the sensitivity for WIMP detection down through the neutrino fog \cite{nufog,nuXeAr}. The first step will be the DarkSide-20k experiment at LNGS; the data taking is intended to start in 2026. The experiment is designed with a goal of an instrumental background $<$0.1~events over a 200~t$\cdot$y exposure for a fiducial mass of 20~t. In parallel, a much smaller detector specifically optimized for the investigation of low-mass dark matter, DarkSide-LowMass, is being considered \cite{dslm}. ARGO will be a multi-hundred tonne detector possibly operated at SNOLAB, having also excellent sensitivity to CNO neutrinos and galactic supernovae \cite{dssn}.



\subsection{Underground Argon}
One of the goals of GADMC is the procurement of large amounts of low-radioactivity UAr as detector target; three projects are in development to ensure this: 
\begin{itemize}
\item Extraction of argon from an underground source (CO$_2$ wells) will be carried out at the Urania plant, in Cortez, CO (US). This is the same source used for the DarkSide-50 detector.
\item UAr will be further chemically purified to detector-grade argon in the Aria facility, in Sardinia (Italy), to remove non-Argon isotopes. Aria will consist of a 350~m cryogenic distillation column, currently being installed. Isotopic distillation with a short version of this column was demonstrated both with nitrogen \cite{aria} and argon isotopes \cite{aria2}. Aria can also be operated in isotopic separation mode to achieve a 10-fold suppression of $^{39}$Ar although at a much reduced throughput \cite{aria2}; this further suppression beyond UAr level is not needed to achieve the physics goals of DarkSide-20k. 
\item Assessing the ultra-low $^{39}$Ar content of the UAr is the the goal of the DArT detector \cite{dart} in construction at the Canfranc Underground Laboratory (LSC) in Spain. 
\end{itemize}

There is a growing interest in the use of ultra-pure UAr outside GADMC, as it has potential broader applications for measuring coherent neutrino scattering in the COHERENT experiment \cite{coherent}, neutrinoless 2$\beta$ decay in the LEGEND-1000 project \cite{legend}, and future modules of the DUNE experiment \cite{dune}; the challenges for its production and characterization are carefully addressed in Refs.~\cite{pnnlworkshop,SnowmassUArFacility}.

\subsection{DarkSide-20k}
\label{secDS20k}

\begin{figure}
\centering
  \includegraphics[height=0.35\textheight]{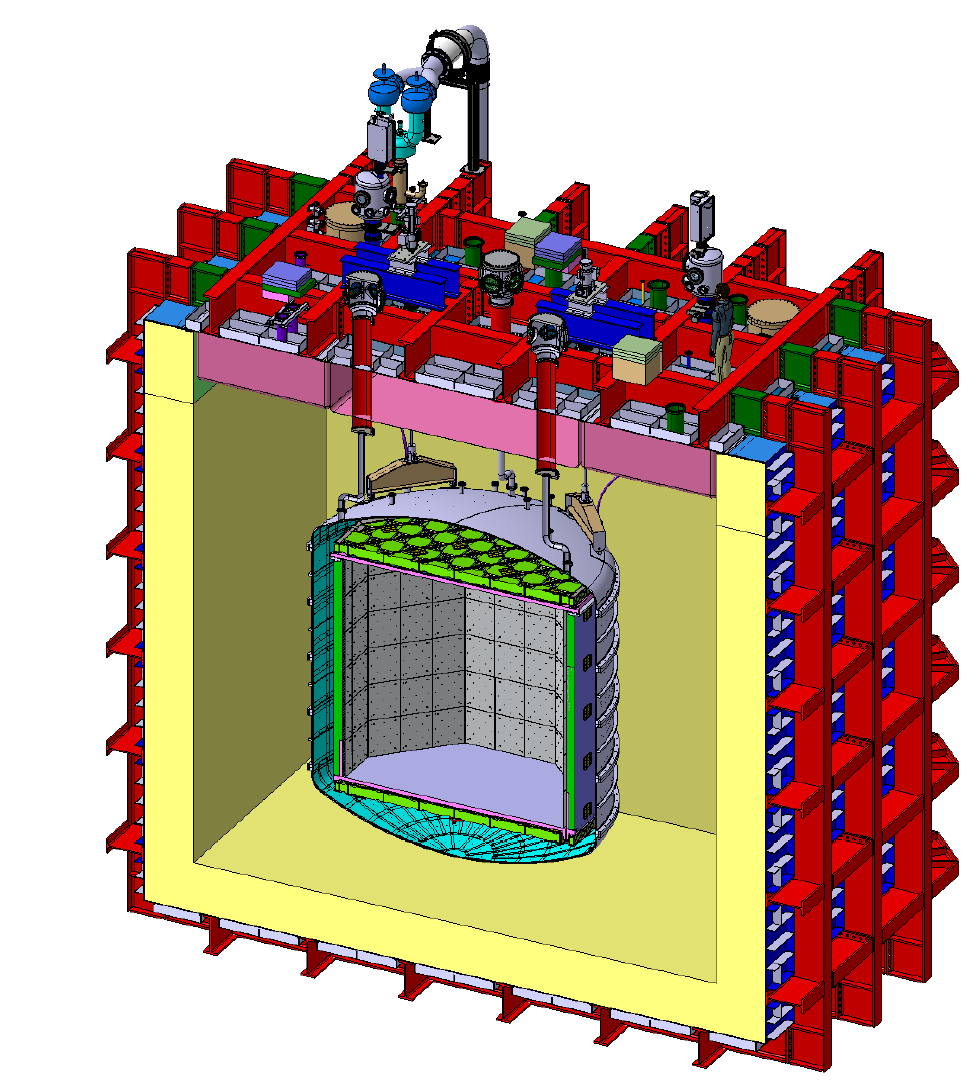}
   \includegraphics[height=0.35\textheight]{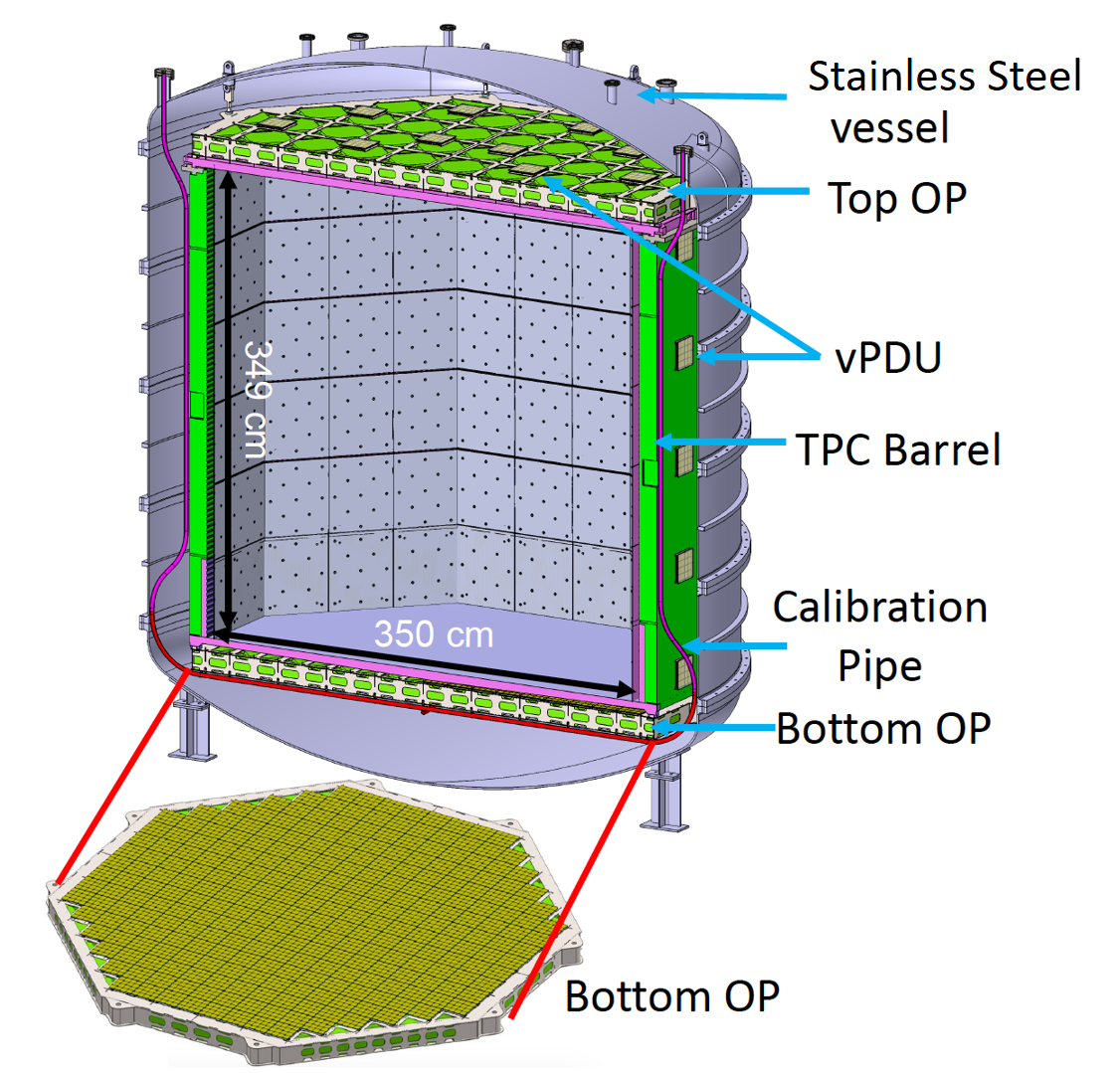} 
  \caption{Cross sections of the cryostat (left) and of the vessel containing the inner veto and TPC (right) of the DarkSide-20k detector. OP stands for Optical Plane and vPDU for veto Photo Detection Unit.}
  \label{dsdesign}
\end{figure}

In DarkSide-20k the core of the apparatus is a dual-phase TPC, serving both as active WIMP target, filled by low-radioactivity UAr \cite{ricap}; a total of 99.2~t of UAr is required, 51.1~t inside the TPC and the rest in the neutron veto. It is planned to produce 120~t of UAr considering contingency. SiPMs in Photo-Detector Modules (PDMs) read the prompt scintillation in the liquid (S1) and delayed electroluminescence in the gas phase (S2). The TPC walls is made of a gadolinium-loaded acrylic vessel (Gd-PMMA); this material is highly efficient at moderating and then capturing neutrons, the capture resulting in the emission of several $\gamma$-rays that allow to tag neutron-induced background events. The detector is housed within a 12-ton vessel, made of stainless steel, immersed in a bath of 700~t of AAr acting as radiation shield and outer veto detector for cosmic background. All the materials used to build the whole detector system are carefully selected for low levels of radioactivity. Figure~\ref{dsdesign} shows cross views of the cryostat and of the inner detector. Table~\ref{tablemass} lists materials, masses and considered cosmogenic isotopes for the main components in the design.

\begin{table}
\centering
\caption{Detector components, materials and masses of the DarkSide-20k detector shown in Fig.~\ref{dsdesign}. Cosmogenically-induced isotopes considered for each material in this work are also indicated; activation in Gd-loaded PMMA has not been analyzed as no hint was found in the radiopurity measurements by $\gamma$ spectroscopy performed for acrylic and Gd$_2$O$_3$ samples.}
\begin{tabular}{l|ccc} \hline
Component & Material & Mass & Induced isotopes \\ \hline
Membrane cryostat & Stainless steel & 224.6~t & See Table \ref{tablecuss} \\
Outer Veto: filling & AAr & 700~t & $^{37}$Ar, $^{39}$Ar, $^{3}$H\\
Inner Veto: vessel & Stainless steel & 12~t  & See Table \ref{tablecuss}\\
TPC: barrel & Gd-loaded PMMA & 11~t & -\\
TPC: grids, frame, brackets & Stainless steel & 1055~kg & See Table \ref{tablecuss}\\
TPC: cables & Copper & 117.8~kg & See Table \ref{tablecuss} \\
Inner Veto+TPC: filling & UAr & 99.2~t & $^{37}$Ar, $^{39}$Ar, $^{3}$H\\
Electronic boards  & Copper & 47.3~kg & See Table \ref{tablecuss}\\ \hline
\end{tabular}
\label{tablemass}
\end{table}

G4DS \cite{g4ds} is a Monte Carlo (MC) simulation framework developed for DarkSide background studies based on GEANT4, providing accurate simulation of light production, propagation, and detection for background and signal events; it has been extensively validated on DarkSide-50 data \cite{g4ds}. For DarkSide-20k, $\gamma$ emissions from the full set of detector components have been simulated to estimate the corresponding background rates in the TPC and in the Veto; activities measured in an extensive material screening campaign based on the combination of different radioassay techniques have been considered. Discrimination techniques based on energy and position of the interactions are implemented to compute the rate in the fiducial volume. As used in \cite{dssn}, preliminary estimates of $\gamma$ background rates point to values around 50~Hz in the TPC and 100~Hz in the neutron Veto, with dominant contribution from PDMs. The $\beta$ contribution of $^{39}$Ar, considering the total active mass of UAr in the TPC (50~tonnes) and in the inner veto (32~tonnes) and the measured activity value in DarkSide-50, yields 36~Hz in the TPC and 26~Hz in the Veto. In this work, cosmogenically induced background shall be compared to these expected rates from radiogenic background from detector material.



\section{Methodology} \label{cal}

One of the most relevant processes in the production of radioactive isotopes in materials is the spallation of nuclei by high energy nucleons; other reactions like fragmentation, induced fission or capture can be important for some nuclei too. On Earth's surface, as the proton to neutron ratio in cosmic rays decreases significantly at energies below the GeV scale because of the absorption of charged particles in the atmosphere, activation by neutrons is usually dominant. Cosmogenic production of radionuclides underground can often be considered negligible, as the flux of cosmic nucleons is suppressed by more than four orders of magnitude for depths of a few tens of meters water equivalent (m.w.e.) \cite{heusser}. Radiogenic neutrons, with fluxes in deep underground facilities that are orders of magnitude lower than that of cosmic neutrons on surface, have energies around a few MeV, too low for spallation processes. 

To quantify the effect of material cosmogenic activation in a particular experiment, the first step is to know the production rates, $R$, of the relevant isotopes induced in the material targets. Then, the produced activity, $A$, can be estimated according to the exposure history to cosmic rays; for instance, considering just a time of exposure $t_{exp}$ followed by a cooling time (time spent underground once shielded from cosmic rays) $t_{cool}$, for an isotope with decay constant $\lambda$, the activity can be evaluated as:
\begin{equation}
A = R [1-\exp(-\lambda t_{exp})] \exp(-\lambda t_{cool}). \label{eqact}
\end{equation}
\noindent Finally, the counting rate generated in the detector by this activity can be computed using G4DS \cite{g4ds}. 

Some direct measurements of production rates at sea level have been carried out for a few materials from the saturation activity, obtained by sensitive screening of samples exposed in well-controlled conditions or by irradiating samples in high flux particle beams. However, in many cases, production rates must be evaluated from the flux of cosmic rays, $\phi$, and the isotope production cross-section, $\sigma$, with both dependent on the particle energy $E$:
\begin{equation}
R=N_t\int\sigma(E)\phi(E)dE \label{eqrate},
\end{equation}
\noindent where $N_t$ is the number of target nuclei. The spread for different calculations of productions rates is usually important, even within a factor 2 (see for instance Tables \ref{tableratesAr} and \ref{tableratesH}). In this work, measured production rates have been used whenever available and dedicated calculations have been performed otherwise.

\subsection{Cosmic ray flux}

An analytic expression for the cosmic neutron spectrum at sea level is presented by Gordon et al in Ref.~\cite{gordon}, deduced by fitting data from a set of measurements for energies above 0.4~MeV; with this parameterization, the integral flux from 10~MeV to 10~GeV is 3.6$\times 10^{-3}$cm$^{-2}$s$^{-1}$ (for New York City). In Ref.~\cite{ziegler}, a similar parametrization is provided as well as correction factors, $f$, to the flux when considering exposure at different locations, as flux depends on the altitude and geomagnetic rigidity. For example, outside LNGS at an altitude of $\sim$1000~m, a correction factor $f=$2.1~\cite{cebrianap} is used. Alternatively, the EXPACS (``EXcel-based Program for calculating Atmospheric Cosmic-ray Spectrum'') program\footnote{EXPACS: \url{https://phits.jaea.go.jp/expacs/}.} could be used to calculate fluxes of nucleons, muons, and other particles for different positions and times in the Earth’s atmosphere; in this way, possible temporal variations of the cosmic rays fluxes are taken into account. Although precise EXPACS calculations are being considered, results presented here are based on the parameterization from Ref.~\cite{gordon} and correction factor from Ref.~\cite{ziegler}.

\subsection{Production cross sections}

Measurements at fixed energies and calculations using different computational codes must be both be taken into account in evaluating $\sigma(E)$. The following have been used in this work:
\begin{itemize}
\item The Experimental Nuclear Reaction Data database (EXFOR, CSISRS in US)~\cite{exfor}, which provides nuclear reaction data and then measured production cross sections \footnote{EXFOR: \url{http://www.nndc.bnl.gov/exfor/exfor.htm}, \url{http://www-nds.iaea.org/exfor/exfor.htm}.}.
\item The Silberberg and Tsao equations presented in Refs.~\cite{tsao1,tsao2,tsao3}, which are semiempirical formulae derived from proton-induced reactions for energies $>$100~MeV and integrated in different codes: COSMO \cite{cosmo}, YIELDX \cite{tsao3} and ACTIVIA \cite{activia}. 
\item The MC simulation of the interaction between projectiles and nuclei, which allows also computation of production cross sections. Many different models and codes have been developed and validated considering the relevant processes. Evaluated libraries of production cross sections have been elaborated, covering different types of reactions or projectiles and different energies, like TENDL (TALYS-based Evaluated Nuclear Data Library)\footnote{\url{https://tendl.web.psi.ch/tendl\_2019/tendl2019.html}} \cite{tendl} (based on the TALYS code, for protons and neutrons with energies up to 200~MeV); JENDL (Japanese Evaluated Nuclear Data Library) \cite{jendl} High Energy File\footnote{JENDL HE library, \url{https://wwwndc.jaea.go.jp/ftpnd/jendl/jendl40he.html}; \url{https://wwwndc.jaea.go.jp/jendl/jendl.html}} (based on the GNASH code, for protons and neutrons from 20~MeV to 3~GeV) is an extension of the JENDL-4.0/HE library including results up to 200~MeV; HEAD-2009 (High Energy Activation Data) \cite{head2009} (for protons and neutrons with higher energies, from 150~MeV up to 1~GeV) uses a selection of models and codes (CEM, CASCADE/INPE, MCNP, etc.).
\end{itemize}

\section{Cosmogenic yields in Copper and Steel}
\label{secCuSt}

The effect on DarkSide-20k of cosmogenic activity in the components made of copper and stainless steel, known to become activated \cite{cebrian,cebrianuniverse}, is analyzed here. 

\subsection{Production rates}
The production rates of the radionuclides typically induced in these materials have been selected from measured and calculated results available in the literature \cite{cebrian,cebrianuniverse}. Estimates using mainly ACTIVIA, GEANT4, and TALYS codes have been made. Saturation activities have been measured with sensitive germanium detectors in samples of copper \cite{coppers,schumann,she} and steel \cite{coppers}, exposed for long times to cosmic rays. In particular, in this work, the production rates from dedicated measurements, using 125~kg of copper provided by Norddeutsche Affinerie (now Aurubis) exposed for 270~days at Gran Sasso and Nironit stainless steel exposed for 314 days, have been considered \cite{coppers}; values are reproduced in Table~\ref{tablecuss}. Among the different products identified in copper, $^{60}$Co has the longest half-life and, unfortunately, there is a significant disagreement on the production rate estimates \cite{cebrian,cebrianuniverse}; the measured value in Ref.~\cite{coppers} is higher than most of the other estimates by a factor of up to a few times. No assessment of $^{60}$Co production in stainless steel is made in Ref.~\cite{coppers}, as the cosmogenically induced activity is shadowed by the intrinsic $^{60}$Co at similar level naturally occurring in typical stainless steel material; for this reason, the rate derived from GEANT4 calculations \cite{mei2016} has been used. Following the half-lives of the different cosmogenic isotopes identified in copper and steel (also shown in Table~\ref{tablecuss}), $^{54}$Mn, $^{57}$Co and $^{60}$Co are in principle the most relevant products.

\subsection{Activity} \label{actCuSt}

To assess the possible effect of the cosmogenic isotopes in these materials for DarkSide-20k, activity $A$ has been evaluated considering the selected production rates at sea level, $t_{cool}=$0 and extreme cases of exposure: $t_{exp}=$1~month, $t_{exp}=$1~year and $t_{exp}=$10~years. It is worth noting that as measured production rates have been taken into account, the deduced activation corresponds to all cosmic ray particles. The final expected activity is obtained from the specific activities derived from the production rates (per mass unit) using Eq. \ref{eqact} and the mass of all the components used in the experimental set-up, which according to the present design of DarkSide-20k are 165.1~kg of copper (mainly from cables and PDMs electronic components) and 226~tons of stainless steel (mainly from cryostat components) plus 12~tonnes from the inner detector. 




Table~\ref{tablecuss} summarizes the total induced activity in copper and stainless steel, respectively, for the relevant isotopes evaluated at the end of the different exposure times; contribution from each individual component is proportional to its mass (see Table~\ref{tablemass}). Following the decay mode of these nuclei, $\gamma$ emissions of the order of 1~MeV will be generated around the active volume by this cosmogenic activation. In the case of copper, even assuming 10~years of exposure, the total activity is at the level of 0.5~Bq. The induced activities are then compared with available measurements from radioassays. For the copper from the Luvata company which is being considered in DarkSide-20k, upper limits of 0.30~mBq/kg of $^{60}$Co and 0.35~mBq/kg of $^{54}$Mn are obtained using a HPGe detector (named GeOroel) in the Canfranc Underground Laboratory. Exposure to cosmic rays of this copper material for a few years can be tolerated since it would contribute a fraction of the upper limit on $^{60}$Co contamination. For all stainless steel components, some cosmogenic activities can be at the level of a few hundreds of Bq, even for just 1~year of exposure; $^{54}$Mn is identified as a potential relevant contributor to the background. Comparing with available measurements from screening, the derived cosmogenic activity of $^{60}$Co is much lower than for instance the one measured for a sample of stainless steel for the DarkSide-20k crysotat, using the same HPGe detector in the Canfranc Underground Laboratory, finding (10.8$\pm$0.9)~mBq/kg of $^{60}$Co. A more stringent requirement of $\sim$1~year of exposure would come by requiring the $^{54}$Mn induced activity being less than the measured one in radio-assay of (1.4$\pm$0.3)~mBq/kg.

\begin{landscape}
\begin{table}
\centering
\caption{Estimates of induced activity in copper and stainless steel components of DarkSide-20k at the end of the exposure to cosmic rays. For each product, the half-life \cite{ddep}, main $\gamma$ emissions and corresponding probabilities are indicated together with the production rates $R$ at sea level considered (from measurements in Ref.~\cite{coppers} except for $^{60}$Co in stainless steel, taken from Ref.~\cite{mei2016}) and the total activity $A$ for the three exposure times considered (1~month, 1~year and 10~years).}
\begin{tabular}{l|cccccccc}
\hline & $^7$Be & $^{46}$Sc & $^{54}$Mn & $^{59}$Fe & $^{56}$Co & $^{57}$Co & $^{58}$Co & $^{60}$Co \\ \hline
T$_{1/2}$ (d) & 53.22 & 83.79 & 312.19 & 44.49 & 77.24 & 271.82 & 70.85 & 1923.95	\\
$\gamma$ emissions (keV) & 477.6  &  889.3, 1120.5  & 834.8  & 1099.3, 1291.6  & 846.8, 1238.3 & 122.1 	& 810.8  & 1173.2, 1332.5 \\
probability (\%) & 10.5 & 99.98, 99.98	& 99.98 & 56.5, 43.2 & 100, 67.6 & 85.6 & 99 & 99.97, 99.99 \\ \hline \hline
{\bf Copper} & & & & & & & & \\
$R$ (atoms/kg/day) &  & 2.18$\pm$0.74	& 8.85$\pm$0.86 & 	18.7$\pm$4.9 & 9.5$\pm$1.2 & 74$\pm$17	& 67.9$\pm$3.7 & 	86.4$\pm$7.8 \\ \hline
$A$ (1~m) (mBq) & & 0.92$\pm$0.31 &	1.09$\pm$0.11 & 13.3$\pm$3.5 & 4.28$\pm$0.54 & 10.4$\pm$2.4 & 33.0$\pm$1.8 & 1.77$\pm$0.16	\\ 	
$A$ (1~y) (mBq) & & 4.0$\pm$1.3	& 9.39$\pm$0.91 & 35.6$\pm$9.3 & 17.5$\pm$2.2 & 86$\pm$20 & 126.1$\pm$6.9 &	20.3$\pm$1.8 \\
$A$ (10~y) (mBq) & & 4.2$\pm$1.4 & 16.9$\pm$1.6 &  35.7$\pm$9.4 & 18.2$\pm$2.3 & 141$\pm$32 & 129.8$\pm$7.1 & 121$\pm$11 \\ \hline \hline
{\bf Stainless Steel} & & & & & & & & \\
$R$ (atoms/kg/day) & 389$\pm$60	& 19.0$\pm$3.5 & 233$\pm$26 & & 20.7$\pm$3.5 & & 51.8$\pm$7.8 &  6.27	\\ \hline
$A$ (1~m) (Bq) & 346$\pm$53 & 11.5$\pm$2.1 & 41.3$\pm$4.6 & & 13.4$\pm$2.3 & & 36.2$\pm$5.5 & 0.19 \\
$A$ (1~y) (Bq) & 1061$\pm$164 & 49.7$\pm$9.2 & 356$\pm$40 & & 54.8$\pm$9.3 & & 138$\pm$21 & 2.1 \\
$A$ (10~y) (Bq) & 1070$\pm$165 & 52.3$\pm$9.6 & 641$\pm$71 & & 56.9$\pm$9.6 & & 142$\pm$21 & 13 \\ \hline \hline
\end{tabular}
\label{tablecuss}
\end{table}
\end{landscape}

\section{Cosmogenic yields in Argon}
\label{secAr}

Argon in the atmosphere contains stable $^{40}$Ar at 99.6\%; cosmogenically produced radioactive isotopes, mainly $^{39}$Ar but also $^{37}$Ar or $^{42}$Ar, can be a significant background if argon obtained from air is used. The concentration of these three isotopes is much reduced in UAr, but the production of cosmogenic radionuclides after extraction must be taken into consideration.

 \subsection{Relevant isotopes}

 $^{39}$Ar is a $\beta^{-}$ emitter with a transition energy of 565~keV and half-life of 269~y \cite{toi}; it is mainly produced by the $^{40}$Ar(n,2n)$^{39}$Ar reaction by cosmic neutrons \cite{saldanha}. The typical activity of $^{39}$Ar in AAr is at the level of $\sim$1~Bq/kg, as measured by WARP \cite{warp}, ArDM \cite{ardm} and DEAP \cite{deap}. In UAr, after a first study on argon from deep underground sources \cite{dsuar}, the measured activity of $^{39}$Ar in the DarkSide-50 detector was (0.73 $\pm$ 0.11)~mBq/kg following a campaign of extracting and purifying argon from deep CO$_2$ wells in Colorado, US; as mentioned in Sec.~\ref{secintro}, this means a reduction of a factor (1.4$\pm$0.2)$\times$10$^3$ relative to the AAr \cite{darkside50}. 

The presence of cosmogenically produced $^{37}$Ar was also detected at the beginning of the run of the DarkSide-50 detector with UAr \cite{darkside50}. It decays 100\% by electron capture to the ground state of the daughter nuclei with a half-life of 35.02~days \cite{ddep}; then, the binding energy of electrons from K-shell (2.8~keV, at 90.21\%) and L-shell (0.20-0.27~keV, at 8.72\%) can be measured as a distinctive signature. The main production channel is the $^{40}$Ar(n,4n)$^{37}$Ar reaction \cite{saldanha}. Underground production in UAr by thermal and epithermal neutron capture is negligible, as for $^{39}$Ar, considering rates as in Ref.~\cite{saldanha} and neutron fluxes at LNGS.
$^{42}$Ar is a pure $\beta^-$ emitter with a 32.9~y half-life and transition energy of 599~keV, generating $^{42}$K, also a $\beta^-$ emitter with half-life of 12.36~h and transition energy of 3525~keV \cite{toi}; this isotope can affect neutrinoless 2$\beta$ experiments using liquid argon as cooling bath and shielding, as shown by the GERDA experiment \cite{gerda42ar} and its specific activity has been studied by ICARUS \cite{icarus}, DBA (92$^{+22}_{-46}$ $\mu$Bq/kg \cite{dba}) and DEAP (40.4$\pm$5.9~$\mu$Bq/kg \cite{deap}). The production rate of $^{42}$Ar in UAr at sea level has been evaluated by GEANT4 simulation as 5.8$\times$10$^{-3}$~atoms/kg/day in Ref.~\cite{zhangmei}; this rate would give from Eq.~\ref{eqact} a saturation activity of 0.07~$\mu$Bq/kg, about three orders of magnitude lower than measured values in AAr. Taking all this into account, the effect of $^{42}$Ar in DarkSide-20k will not be considered here although a specific study to quantify radiogenic and cosmogenic production in the Earth's crust is underway\footnote{\url{https://indico.sanfordlab.org/event/29/contributions/487/}}.

$^3$H is a pure $\beta^-$ emitter with transition energy of 18.6~keV and a long half-life of 12.3~y~\cite{ddep}. The quantification of its cosmogenic production is not easy by calculations ($^3$H can be generated by different reaction channels) nor experimentally (the $\beta$ emissions are hard to disentangle from other background contributions). Estimates of the $^3$H production rate in several dark matter targets were attempted in Ref.~\cite{tritiumpaper}; the rate has been measured for germanium from EDELWEISS \cite{edelweisscos} and CDMSlite \cite{cdmslitecos} data and for silicon and NaI(Tl) from neutron irradiation \cite{saldanhasi,saldanhaNaI}. The possible presence of $^3$H has been observed also in NaI(Tl) crystals by the ANAIS \cite{cebrianjcap,naiepjc2019} and COSINE experiments \cite{cosinecos,cosinebkg}. In principle, purification systems for LAr may remove all non-argon radionuclides and $^3$H should not be a problem for DarkSide. This was also assumed for liquid xenon, but $^3$H was considered as a possible explanation for the excess of electronic recoil events observed in the XENON1T experiment below 7~keV \cite{xenon1texcess,robinson}, which was not observed in XENONnT \cite{xenonnt}. Activated $^3$H is separated from argon with SAES Getters \cite{meikrantzTritiumProcessApplications1995} and will be removed {\it in situ} while the UAr recirculates.

Other radioisotopes with half-lives longer than 10~days like $^7$Be, $^{10}$Be, $^{14}$C, $^{22}$Na, $^{26}$Al, $^{32}$P, $^{33}$P, $^{32}$Si, $^{35}$S, $^{36}$Cl, $^{40}$K and $^{41}$Ca are also produced in argon, as shown using the COSMO code. The production rates of these isotopes at sea level from fast neutrons, high energy muons and protons have been evaluated by GEANT4 simulation in Ref.~\cite{zhangmei}. Assuming an efficient purification of non-noble isotopes, they will not be considered in this study.

\subsection{Production rates}
\label{secpr}

The production rates of $^{37}$Ar and $^{39}$Ar from cosmic neutrons at sea level were measured for the first time through controlled irradiation at Los Alamos Neutron Science Center (LANSCE) with a neutron beam resembling the cosmic neutron spectrum and later direct counting with sensitive proportional counters at Pacific Northwest National Laboratory (PNNL) \cite{saldanha}. In addition, the study of other production mechanisms due to muon capture, cosmic protons and high energy $\gamma$ rays at the Earth's surface was made using available cross sections to compute total production rates at sea level. 
The production rates obtained in Ref.~\cite{saldanha} for UAr are reproduced in Table~\ref{tableAAr} as they will be used to evaluate the induced activity in DarkSide-20k. The production rates of both $^{37}$Ar and $^{39}$Ar at sea level were also evaluated by GEANT4 simulation in Ref.~\cite{zhangmei}.

 The UAr to be used in DarkSide-20k is extracted in Colorado, at a quite high altitude, so the corresponding correction factors $f$ to the cosmic ray flux at sea level must be taken into consideration. In Ref.~\cite{ziegler}, high values of $f$ are reported for neutrons at Colorado locations: 4.11 and 12.86 for Denver (at 1609~m) and Leadville (at 3109~m), respectively. 
These correction factors $f$ have been adjusted to the altitude at the Urania facilities (at 2164~m), assuming that the ratio of $f$ for different altitudes is the same as the ratio of cosmic flux intensities. As described in Ref.~\cite{ziegler}, the intensities $I_1$ and $I_2$ at two different altitudes $A_1$ and $A_2$ (converted to g/cm$^2$) are related as:
\begin{equation}
    I_2=I_1 \exp[(A_1-A_2)/L], \label{eqI}
\end{equation}
\noindent being $L$ the absorption length for the cosmic ray particles. Calculations for the cosmic neutron flux correction factor are summarized in Table~\ref{tablef}, using $L=$136~g/cm$^2$; the final result for Urania is the average between those from Denver and Leadville data, $f=$6.43. For cosmic protons and muons, the correction factors have been obtained just from Eq.~\ref{eqI} considering the corresponding absorption lengths ($L=110$~g/cm$^2$ for protons and $L=261$~g/cm$^2$ for muons \cite{ziegler}); the results are $f=8.67$ for protons and $f=2.48$ for muons.

\begin{table}
\centering
\caption{Calculation of the correction factor $f$ to be applied to the cosmic neutron flux at sea level (in New York) for the location of the Urania facilities in Colorado. The relative intensities $I$ are derived from Eq.~\ref{eqI}. The final factor for Urania is the average between the deduced ones from Denver and Leadville data.}
\begin{tabular}{l|ccccc} \hline
Location &	$H$ &	$A$  &	$f$  & Relative $I$  & Deduced $f$  \\
 & (m)  & (g/cm$^2$) & from Ref.~\cite{ziegler} & to Urania & for Urania \\ \hline
Denver &	1609  &	852.3 &	4.11 &	0.659 &	6.24 \\ 
Leadville &	3109  &	705.2 &	12.86 &		1.942 &	6.62 \\ \hline 
Urania & 	2164  &	795.5  & & & 6.43 \\ \hline 
\end{tabular}
\label{tablef}
\end{table}		

\begin{figure}
\centering
  \includegraphics[height=.35\textheight]{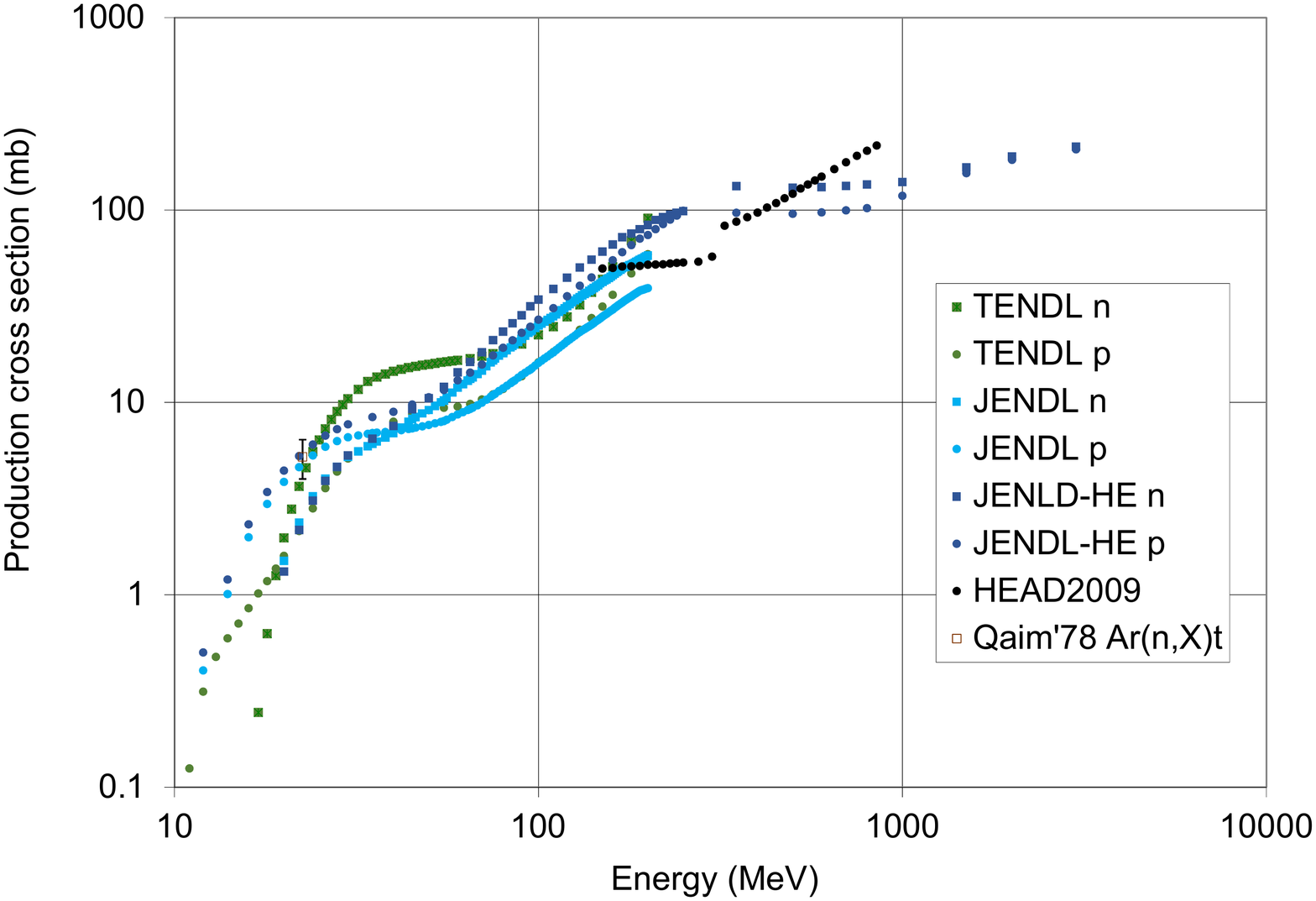}
  \includegraphics[height=.35\textheight]{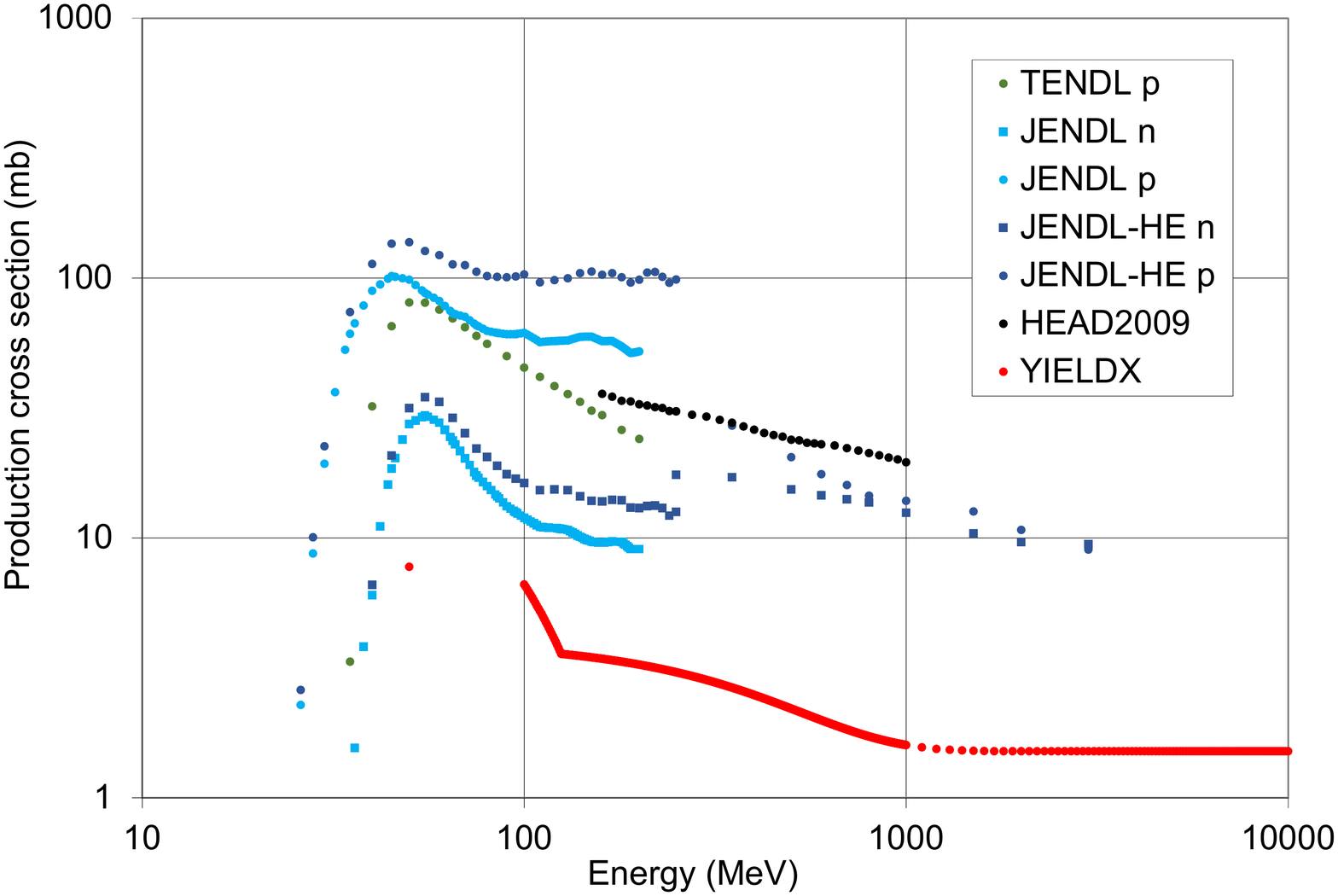} 
  \includegraphics[height=.35\textheight]{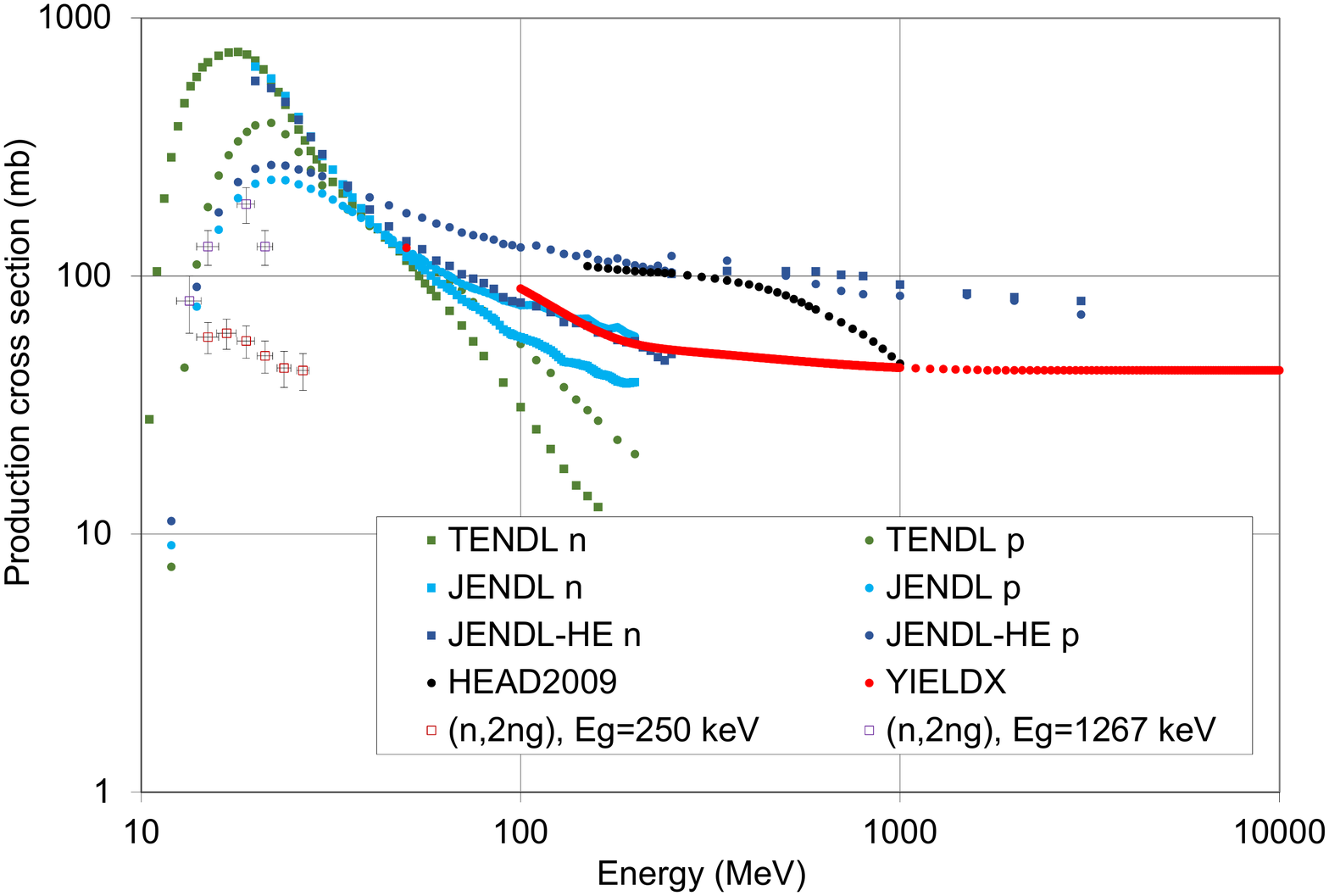}
  \caption{Production cross sections of $^{3}$H (top), $^{37}$Ar (middle) and $^{39}$Ar (bottom) in $^{40}$Ar by nucleons vs energy taken from different sources.}
  \label{excfuc}
\end{figure}

Following Eq.~\ref{eqrate}, a calculation of the production rates of relevant isotopes in argon (assuming 100\% $^{40}$Ar) by cosmic neutrons from Ref.~\cite{gordon} has been made considering a selection of excitation functions from libraries and YIELDX calculations. Figure \ref{excfuc} shows our compilation of production cross sections of $^{3}$H, $^{37}$Ar and $^{39}$Ar by nucleons. For $^{39}$Ar, although no experimental data at EXFOR was found for the total production cross section, there are results for partial (n,2n$\gamma$) reactions in natural argon at 1-30 MeV taken from Ref.~\cite{macmullin}. For $^{3}$H, an irradiation experiment with neutrons having an energy spectrum peaked at 22.5 MeV measured the corresponding production cross section~\cite{qaim}.

A mismatch between cross section data from different libraries is observed.  Several descriptions of the cross sections, even from different libraries below and above a particular energy cut, have been considered to estimate the corresponding uncertainty; the obtained maximum and minimum rates define an interval, whose central value and half width have been considered as the final result and its uncertainty for the evaluation of the production rates. Table~\ref{tableratesAr} presents the obtained results for $^{37}$Ar and $^{39}$Ar, together with the measured production rate for fast neutrons and different calculations from Refs.~\cite{saldanha,zhangmei}. The production rate of $^{39}$Ar derived here is fully compatible with the measured value (and with several of the calculations in Ref.~\cite{saldanha}). The production rate of $^{37}$Ar is a factor 2 higher than the measured one, but lower than the GEANT4 estimate in Ref.~\cite{zhangmei}. For calculating the final activity yields of $^{37}$Ar and $^{39}$Ar, the values of the total production rates obtained in Ref.~\cite{saldanha} will be used; but this comparison can be useful to assess the reliability of the production rates of isotopes estimated only from calculations, like $^3$H in argon.

The production rate of $^3$H in argon was calculated, as for other targets, using different codes like TALYS \cite{mei} and GEANT4 and ACTIVIA \cite{mei2016}. It was also computed in Ref.~\cite{tritiumpaper} using a similar approach as used in this work from a selection of excitation functions considering the TENDL and HEAD2009 libraries. The results ranged from 115.1 to 177.2~atoms/kg/day and the approach was cross-checked against experimental data for NaI and germanium, reproducing properly measured production rates \cite{edelweisscos,cdmslitecos,saldanhaNaI}. We add to the analysis new data included in the JENDL-HE library which gives a  production rate of 221.6~atoms/kg/day. We combine the results in Ref.~\cite{tritiumpaper} with this latter one to estimate a central value and uncertainty for the production of $^3$H as (168$\pm$53)~atoms/kg/day. 
It must be noted that this value gives only production by neutrons; assuming equal flux and cross sections of protons and neutrons above 1 GeV, it is estimated that protons would increase the rate by 10\% at most \cite{tritiumpaper} and is thus neglected in the following. Table~\ref{tableratesH} compares the production rate estimated in this work with all the available ones for $^3$H production in argon taken from the literature following different approaches; an important dispersion of values is found.

\begin{landscape}
\begin{table}
\centering
\caption{Calculations of the production rates $R$ of $^{37}$Ar and $^{39}$Ar in Ar at sea level from this work considering different descriptions of the excitation functions below (LE) and above (HE) a cut energy value; the final estimated rates are given by the ranges defined between the maximum and minimum obtained rates (see text). Different calculations from the literature (considering the same cosmic neutron spectrum from Ref.~\cite{gordon}) and the measured value for fast neutrons from Ref.~\cite{saldanha} are also shown for comparison.}
\begin{tabular}{lcc|lcc} \hline
& & $^{37}$Ar & & & $^{39}$Ar  \\ \hline \hline
This work: & Cut  & $R$  & This work & Cut  & $R$  \\ 
LE+HE & (MeV) & (atoms/kg/day) & LE+HE & (MeV) & (atoms/kg/day) \\ \hline
TENDL(p)+HEAD2009 &	150	& 153.6 & TENDL+HEAD2009 &	150	& 726.4 \\
TENDL(p)+YIELDX	& 100	& 93.5 & TENDL+YIELDX &	100 & 	697.1 \\
TENDL(p)+YIELDX	& 200 & 122.7 & TENDL+YIELDX &	200	& 646.0 \\
JENDL-HE(n)	& 30 & 	63.9 & TENDL+JENDL-HE(n) &	20	& 804.3 \\ \hline
Estimated rate in this work  & & 109$\pm$45 & Estimated rate in this work & & 725$\pm$79 \\ \hline \hline
Not used for estimation: &&&&& \\
Measurement \cite{saldanha} & & 51.0$\pm$7.4 & & & 759$\pm$128 \\
ACTIVIA \cite{saldanha} & & 17.9$\pm$2.2 & & & 200$\pm$25 \\
MENDL-2P \cite{saldanha} & & 155$\pm$19 & & & 188$\pm$24 \\
TALYS \cite{saldanha} & & 76.8$\pm$9.6 & & & 753$\pm$94 \\
INCL++ (ABLA07) \cite{saldanha} & & 79.3$\pm$9.9 & & & 832$\pm$104 \\
 & & & TENDL-2015 \cite{saldanha} & & 726$\pm$91 \\
GEANT4 \cite{zhangmei} & & 176 &  & & 858 \\
 \hline
\end{tabular}
\label{tableratesAr}
\end{table}
\end{landscape}

\begin{table}
\centering
\caption{Production rate $R$ of $^3$H in Ar at sea level from this work and from different calculations from the literature.}
\begin{tabular}{lc}
\hline
& $R$ (atoms/kg/day) \\ \hline \hline
TENDL & 115.1 \\
HEAD2009 & 177.2 \\
JENDL-HE & 221.6 \\ \hline
Estimated rate in this work & 168$\pm$53 \\ \hline \hline
Not used for estimation: & \\
TALYS \cite{mei} & 44.4 \\
GEANT4 \cite{mei2016} & 84.9  \\
ACTIVIA \cite{mei2016} & 82.9 \\ \hline
\end{tabular}
\label{tableratesH}
\end{table}

\subsection{Activity} \label{actAr}

The possible activity yields of relevant cosmogenic isotopes in Ar have been analyzed for the DarkSide-20k detector considering Ar extraction, storage and transportation and taking into account different cosmic ray components. For $^{37}$Ar and $^{39}$Ar, the production rates at sea level precisely determined with the LANSCE neutron beam and the estimates for muons, protons and cosmic $\gamma$ rays \cite{saldanha} have been considered, while for $^3$H the production rate estimated in this work has been assumed.

The UAr extracted at the Urania plant will be shipped firstly to the Aria facility for purification and then to LNGS for storage and final operation. The current baseline design is to ship the UAr in high-pressure gas cylinders that are organized into skids capable of containing $\sim$2~t of UAr each. The following steps are foreseen:

\begin{enumerate}
\item Storage of UAr at Urania: three skids will be filled before starting transportation. Considering the time required to fill one, exposures of 8, 16 and 24~days have been assumed for each skid. At the Urania site, the UAr will always be on surface while being processed and once in the skids. The correction factors to the sea level fluxes of cosmic neutrons, protons and muons evaluated for Urania location in Colorado (see Sec.~\ref{secpr}) have been included in this step.
\item Trip from Urania to a shipping port: a container with the three skids will transport the UAr from Urania to Houston, TX (USA), by road. An exposure of 7~days has been considered. To take into account the different altitude during the trip, the average between the maximal (from Urania altitude) and minimal (at sea level) expected activity has been calculated. 
\item Trip overseas to Europe: 60~days of exposure at sea level have been conservatively assumed for the trip by boat from Houston to Cagliari. An additional exposure of 7~days is foreseen for custom clearing and the trip from Cagliari to the Aria location.

In total, 16 months are required for completing the extraction and transportation of all the necessary UAr from Urania to Italy.
\item Processing and storage of UAr at Aria: once in Sardinia, the skids will be stored near Aria and the UAr will be accumulated for processing. At a purification rate of 1~ton per day, an expected exposure of 60~days to process two batches of 60~t each has been considered.   
Underground storage at a depth of at least some tens of m.w.e. would be ideal and it is assumed here but, if not possible, an almost linear increase of 2.6~$\mu$Bq/kg in the activity of $^{39}$Ar is estimated per month of additional exposure at sea level.

\item Trip from Aria to LNGS: 10~days of exposure at sea level have been considered for this trip by sea. It is expected to ship 12~t at a time using six skids.
\item Storage at LNGS: skids will be stored underground as they arrive. 
\end{enumerate}


Under these assumptions, the total time from the beginning of production at Urania to the end of processing at Aria is 614~days. 

Taking into account this exposure history, the induced activity by each cosmic ray component has been computed for each exposure step (at Urania, trip in US, overseas, at Aria and trip in Italy) from Eq.~\ref{eqact}. Tables \ref{tableAAr} and \ref{tableAH} show separately each contribution for $^{39}$Ar and $^{37}$Ar and for $^3$H, respectively. Contributions from different cosmic ray components are assumed to be independent to derive uncertainties in total activity. The decrease of the activities induced at each step during the rest of the whole process is negligible for $^{39}$Ar and small for $^3$H, due to their long half-lives, but extremely relevant for $^{37}$Ar; it is accounted for in the final activities reported in Tables \ref{tableAAr} and \ref{tableAH}.

For both $^{39}$Ar and $^{37}$Ar, cosmogenic neutrons are responsible of most of the induced activity. Under the assumed conditions, the relative contributions to the final $^{39}$Ar activity of each exposure step are the following: Urania, 34.4\%; US trip, 9.0\%; overseas trip, 27.7\%; at Aria, 24.8\%; and Italy trip, 4.1\%. The exposure at Urania gives the largest contribution, followed by that of the overseas trip and at Aria. For $^{37}$Ar, having a much shorter half-life, the last exposure during the Italy trip is dominant, producing 55\% of the final activity. Concerning $^3$H, the final activity in Table~\ref{tableAH} would apply if no purification procedures were considered; however, if a 100\% efficient removal of $^3$H was achieved in Aria, only the activity in the last step for exposure in Italy would be produced. Table~\ref{tableSummary} summarizes the expected activities once all the UAr is at LNGS. From values in Table \ref{tableAAr}, the final estimated activity of $^{39}$Ar is (20.7$\pm$2.8)~$\mu$Bq/kg; this equals 2.8\% of measured activity in DarkSide-50. For $^{37}$Ar, the effect of cooling is very important and the expected activity when all the UAr is at LNGS is (103$\pm$14)~$\mu$Bq/kg. From values in Table~\ref{tableAH} for $^{3}$H, an activity of (2.97$\pm$0.94)~$\mu$Bq/kg is expected at that time considering only activation after ideal purification in Aria; with no purification, it would be around 25 times higher. 

Uncertainties quoted for activities in Tables \ref{tableAAr} and \ref{tableAH} come from those of production rates, reproduced in the same tables. Concerning the correction factors of sea level cosmic ray fluxes for exposure at Urania, it has been checked that considering a description different to that applied in Sec.~\ref{secpr} produces very similar results; correction factors computed from EXPACS spectra in the energy range relevant for activation (1~MeV to 10~GeV) are $f=6.09$ for neutrons, $f=7.60$ for protons and $f=1.61$ for muons, giving a small decrease in the final activities: 1.0\% for $^{39}$Ar, no change for $^{37}$Ar and 1.5\% for $^3$H with no purification. On the other hand, unexpected events can produce relevant deviations from the baseline exposure conditions and their effect on the activation yields has been assessed. Doubling the exposure at Urania would increase the final $^{39}$Ar activity from (20.7$\pm$2.8)~$\mu$Bq/kg to (27.7$\pm$3.9)~$\mu$Bq/kg, which would be 3.8\% of the DarkSide-50 activity. Exposure at Aria has been evaluated for the moment considering just the processing time, but activation produced in the periods before and after the processing should be added if storage is made above ground; to produce an additional 10\% of the measured activity in DarkSide-50 (which was determined with an uncertainty of 14\%), 28~months of additional exposure would be required, which is well above the period of 16 months needed for the extraction of the whole amount of UAr needed. It can be concluded that there is enough contingency in the plan for production, storage and shipping of the UAr so that cosmogenic $^{39}$Ar activity does not endanger DarkSide-20k sensitivity.

\section{Expected counting rates in DarkSide-20k}
\label{secRates}


The rates from the estimated cosmogenic activity of products in UAr, under the assumed exposure conditions, are also shown in Table~\ref{tableSummary}. Induced $^{39}$Ar due to the whole exposure from Urania to LNGS would add a rate of (1.03$\pm$0.14)~Hz for the TPC. The contribution of $^{37}$Ar (being (5.15$\pm$0.68)~Hz if data taking started just immediately after the arrival of all the UAr at LNGS) will decay very quickly. Comparing these numbers with the total $\beta$ and $\gamma$ rates presented in Sec.~\ref{secDS20k}, it can be concluded that cosmogenic activity does not produce a problematic increase of the TPC and Veto rates.

\section{Conclusions}
\label{con}

For DarkSide-20k, material cosmogenic activation is a source of $\beta/\gamma$ background and it has been quantified for LAr and other materials used in large amounts from realistic exposure conditions in order to assess the contribution to the counting rates and decide if additional exposure restrictions are necessary. The main results are summarized in Table~\ref{tableSummary}.

\begin{landscape}	
\begin{table}
\centering
\caption{Calculation of the expected induced activity in kg$^{-1}$ d$^{-1}$ of $^{39}$Ar and $^{37}$Ar in the UAr of the DarkSide-20k detector, for the assumed production rates $R$ and exposure times (see text). Different columns and rows show separate contributions by cosmic ray components and exposure steps, respectively; relative contributions of each component to the total activity are also quoted. Row labelled as "Final" presents the sum of final activities from all exposure steps including properly their decays.}
\begin{tabular}{lccccc}
\hline
{\bf $^{39}$Ar} & Neutrons & Muons & Protons & $\gamma$ rays & Total  \\ \hline
$R$ (atoms/kg/day) \cite{saldanha} & 759$\pm$128 & 172$\pm$26 & 3.6$\pm$2.2 & 112.8$\pm$20.9  \\ \hline
Urania & 0.551$\pm$0.093 & 0.0483$\pm$0.0073 & 0.0035$\pm$0.0022 &  0.0127$\pm$0.0024 &  0.616$\pm$0.093 	\\
US & 0.139$\pm$0.024 & 0.0148$\pm$0.0022 & 0.0009$\pm$0.0005 & 0.0056$\pm$0.0010 & 0.161$\pm$0.024 	\\
Overseas & 0.359$\pm$0.061 & 0.081$\pm$0.012 & 0.0017$\pm$0.0010 & 0.053$\pm$0.010 & 0.495$\pm$0.063  \\	
Aria & 0.321$\pm$0.054 & 0.073$\pm$0.011 & 0.0015$\pm$0.0009 & 0.048$\pm$0.0088 &  0.444$\pm$0.056 	\\
Italy & 0.0536$\pm$0.0090  & 0.0121$\pm$0.0018  & 0.0003$\pm$0.0002  & 0.0080$\pm$0.0015  & 0.0739$\pm$0.0093  	\\  \hline
Final & 1.42$\pm$0.24 & 0.229$\pm$0.035 & 0.0078$\pm$0.0048 & 0.127$\pm$0.024 &  1.79$\pm$0.24 \\ 
(\%) & 79.6 & 12.8 & 0.4 & 7.1 & \\ \hline \hline
{\bf $^{37}$Ar} &Neutrons & Thermal neutrons & Protons & $\gamma$ rays & Total   \\ \hline
$R$ (atoms/kg/day) \cite{saldanha} & 51$\pm$7.4 & 0.9$\pm$0.3 & 1.3$\pm$0.4 & 3.5$\pm$0.7 \\ \hline
Urania & 87$\pm$13 &  2.99$\pm$0.92 &  0.93$\pm$0.19 &  0.239$\pm$0.080 &  91$\pm$13 \\
US & 24.5$\pm$3.6 &  0.81$\pm$0.25 & 0.453$\pm$0.091 & 0.116$\pm$0.039 & 25.9$\pm$3.6 \\
Overseas & 37.5$\pm$5.4 & 0.95$\pm$0.29 & 2.57$\pm$0.51 & 0.66$\pm$0.22 & 41.7$\pm$5.5 \\
Aria & 35.5$\pm$5.1 & 0.90$\pm$0.28 & 2.43$\pm$0.49 & 0.63$\pm$0.21 & 39.4$\pm$5.2 \\
Italy & 9.2$\pm$1.3 & 0.234$\pm$0.072 & 0.63$\pm$0.13 &  0.162$\pm$0.054 &  10.2$\pm$1.3 \\ \hline
Final & 8.0$\pm$1.2 & 0.209$\pm$0.064 & 0.52$\pm$0.10 & 0.135$\pm$0.045 & 8.9$\pm$1.2\\
(\%) & 90.3 & 2.3 &  5.9 & 1.5 & \\ \hline
\hline
\end{tabular}
\label{tableAAr}
\end{table}	
\end{landscape}	

\begin{table}
\centering
\caption{Calculation of the expected induced activity in kg$^{-1}$ d$^{-1}$ of $^{3}$H by cosmic neutrons in the UAr of the DarkSide-20k detector, for the production rate $R$ estimated in this work and the assumed exposure times (see text), considering no purification procedure.  Different rows show separate contributions by exposure steps. Row labelled as ``Final'' presents the sum of final activities from all exposure steps including properly their decays.}
\begin{tabular}{lc}
\hline
{\bf $^{3}$H} &  \\ \hline
$R$ (atoms/kg/day) & 168$\pm$53  \\ \hline
Urania & 2.66$\pm$0.84	\\
US & 0.67$\pm$0.21 	\\
Overseas & 1.73$\pm$0.54  \\
Aria & 1.55$\pm$0.49 \\
Italy & 0.259$\pm$0.082  \\ \hline
Final & 6.5$\pm$2.1 \\ \hline
\end{tabular}
\label{tableAH}
\end{table}

\begin{landscape}
\begin{table}
\centering
\caption{Summary table of estimated activation in DarkSide-20k including isotope, material, calculation details, overall activity and counting rates in TPC and inner veto. The most relevant channel for each isotope are shown in the third column (although other ones are included). All reported activity and rate values correspond to the moment when the materials are brought underground. For $^{3}$H, row (1) and (2) assume no purification and ideal purification at Aria, respectively.}
\begin{tabular}{lcccccc}
\hline
Isotope & Material & Most relevant channel & Calculation & Activity  & TPC rate  & Veto rate  \\
& & & & ($\mu$Bq/kg) & (Hz) & (Hz) \\ \hline
$^{39}$Ar & UAr & $^{40}$Ar(n,2n)$^{39}$Ar & Production rates from \cite{saldanha} & 20.7$\pm$2.8 & 1.03$\pm$0.14 & 0.662$\pm$0.090 \\ 
$^{37}$Ar & UAr & $^{40}$Ar(n,4n)$^{37}$Ar & Production rates from \cite{saldanha} & 103$\pm$14 & 5.15$\pm$0.68 &  3.30$\pm$0.43 \\ 
$^{3}$H (1) &  UAr & $^{40}$Ar(n,*)$^{3}$H & $\sigma$(E) in Fig.~\ref{excfuc}+Gordon spectrum & 76$\pm$24 & 3.8$\pm$1.2& 2.42$\pm$0.76 \\
$^{3}$H (2) &  UAr & $^{40}$Ar(n,*)$^{3}$H  & $\sigma$(E) in Fig.~\ref{excfuc}+Gordon spectrum & 2.97$\pm$0.94 & 0.148$\pm$0.047 & 0.095$\pm$0.030 \\ \hline 	
\end{tabular}
\label{tableSummary}
\end{table}	
\end{landscape}

For copper and stainless steel components, activation yields of isotopes with relevant half-lives (like $^{54}$Mn, $^{57}$Co and $^{60}$Co) have been computed from the measured production rates at sea level at Ref.~\cite{coppers}. In copper, even for 10~y of exposure to cosmic rays, estimated activities are below 0.5~Bq. In stainless steel, hundreds of Bq are expected for some isotopes for just 1~y exposure; the contribution to the counting rate of ER-like events in the TPC from $^{54}$Mn activity induced in steel components has been found to be negligible in comparison to the estimated total rate from $\beta/\gamma$ backgrounds. This avoids restricting the surface residency time.

A total of 120~t of UAr depleted in $^{39}$Ar must be extracted and processed for filling the TPC and inner veto of DarkSide-20k. The possible induced activity on surface, from the extraction at Urania to the storage at LNGS, has been analyzed not only for $^{39}$Ar but also for $^{37}$Ar and $^{3}$H. Production rates from Ref.~\cite{saldanha}, based on a neutron irradiation experiment, have been considered for the Ar isotopes while for $^{3}$H an estimate of the production rate by cosmic neutrons made in this work obtaining (168$\pm$53)~atoms/kg/day has been used. The estimated cosmogenic activity of $^{39}$Ar when all the UAr arrives to LNGS, (20.7$\pm$2.8)~$\mu$Bq/kg for the assumed exposure history, is considered acceptable as it is just 2.8\% of the residual activity measured in DarkSide-50 for UAr of the same source and would add $\sim$1~Hz to the counting rate of the TPC. The quantified effect of some uncertain steps in the procedure of UAr production shows that there is enough contingency. Contributions from the induced activity of $^{37}$Ar and $^{3}$H are not problematic thanks to short half-life and purification, respectively. The results of this study of the cosmogenic activation of UAr will be useful to set exposure limitations for the procurement of the large amounts of radiopure UAr necessary in future LAr projects.

\section*{Acknowledgements}
This report is based upon work supported by FSC 2014-2020 - Patto per lo Sviluppo, Regione Sardegna, Italy, the U. S. National Science Foundation (NSF) (Grants No. PHY-0919363, No. PHY-1004054, No. PHY-1004072, No. PHY-1242585, No. PHY-1314483, No. PHY- 1314507, associated collaborative grants, No. PHY-1211308, No. PHY-1314501, and No. PHY-1455351, as well as Major Research Instrumentation Grant No. MRI-1429544), the Italian Istituto Nazionale di Fisica Nucleare (Grants from Italian Ministero dell’Istruzione, Università, e Ricerca Progetto Premiale 2013 and Commissione Scientific Nazionale II), the Natural Sciences and Engineering Research Council of Canada, SNOLAB, and the Arthur B. McDonald Canadian Astroparticle Physics Research Institute. We acknowledge the financial support by LabEx UnivEarthS (ANR-10-LABX-0023 and ANR18-IDEX-0001), the São Paulo Research Foundation (Grant FAPESP-2017/26238-4), Chinese Academy of Sciences (113111KYSB20210030) and National Natural Science Foundation of China (12020101004). The authors were also supported by the Spanish Ministry of Science and Innovation (MICINN) through the grant PID2019-109374GBI00, the “Atraccion de Talento” Grant 2018-T2/ TIC-10494, the Polish NCN, Grant No. UMO- 2019/ 33/ B/ ST2/ 02884, the Polish Ministry of Science and Higher Education, MNi-SW, grant number 6811/IA/SP/2018, the International Research Agenda Programme AstroCeNT, Grant No. MAB-/2018/7, funded by the Foundation for Polish Science from the European Regional Development Fund, the European Union’s Horizon 2020 research and innovation program under grant agreement No 952480 (DarkWave), the Science and Technology Facilities Council, part of the United Kingdom Research and Innovation, and The Royal Society (United Kingdom), and IN2P3-COPIN consortium (Grant No. 20-152). I.F.M.A is supported in part by Conselho Nacional de Desenvolvimento Científico e Tecnológico (CNPq). We also wish to acknowledge the support from Pacific Northwest National Laboratory, which is operated by Battelle for the U.S. Department of Energy under Contract No. DE--AC05-76RL01830. This research was supported by the Fermi National Accelerator Laboratory (Fermilab), a U.S. Department of Energy, Office of Science, HEP User Facility. Fermilab is managed by Fermi Research Alliance, LLC - (FRA), acting under Contract No. DE-AC02-07CH11359.

\end{document}